\documentclass{ws-procs975x65}

\begin{document}

\title{THE "APPROACH UNIFYING SPIN AND CHARGES" PREDICTS THE FOURTH FAMILY AND A STABLE
FAMILY FORMING THE DARK MATTER CLUSTERS}
\author{N. S. MANKO\v C BOR\v STNIK$^*$}

\address{Department of Physics, Faculty of Mathematics and Physics, University of Ljubljana,\\
Jadranska 19, 1000 Ljubljana, Slovenia\\
$^*$E-mail: norma.mankoc@fmf.uni-lj.si}

\begin{abstract}
The {\it Approach unifying spin and charges}~\cite{n92n93n95n01n07,pn06,%
hn00hn02hn03hn06hn07hn08,gmdn07}, assuming that all the internal degrees of freedom---the
spin, all the charges and the families---originate in $d> (1+ 3)$ in only two kinds of spins 
(the Dirac one and the only one existing beside the Dirac one and anticommuting with the Dirac one), 
is offering a new way in understanding the appearance of the families and the charges 
(in the case of charges the similarity with the Kaluza-Klein-like theories must be emphasized).
A simple starting action in $d>(1+3)$ for gauge fields (the vielbeins and the two kinds of the 
spin connections) and a spinor (which carries only two kinds of spins and  
interacts with the corresponding gauge fields) manifests after particular breaks of the 
starting symmetry the massless {\it four}  (rather than three) families 
with  the properties as  assumed by  the {\it Standard model} for the 
three known families, 
and the additional four massive families. The lowest of these additional four families is stable. 
A part of the starting action contributes, together with the vielbeins, in the break of 
the electroweak symmetry manifesting in  $d=(1+3)$ 
the Yukawa couplings (determining the mixing matrices and the masses of the lower four families 
of fermions and influencing the properties of the higher four families) and the scalar field, which 
determines the masses of the gauge fields. The fourth family might be seen at the LHC, while the stable 
fifth family might be what is observed as the dark matter. 
\end{abstract}

\keywords{Origin of families and charges; origin of Yukawa couplings; origin 
of gauge fields; prediction of the fourth family; prediction of stable 
fifth family forming the dark matter; two kinds of the Clifford objects.} 
\bodymatter
\section{Introduction}
\label{introduction}
The {\it Standard model of the electroweak and colour interactions} (extended by the right handed  
neutrinos), fiting with around 25 assumptions and parameters all the existing experimental 
data, leaves  unanswered many open questions, among which are the questions   
about the origin of the charges ($U(1), SU(2), SU(3)$), of the families, and correspondingly 
of the Yukawa couplings of quarks and leptons and of the Higgs mechanism. Answering the question about 
the origin of families and their masses seems to me the most promising way leading beyond the 
today's knowledge about the elementary fermionic and bosonic fields. 

The {\it Approach unifying spins and 
charges}~\cite{n92n93n95n01n07,pn06,hn00hn02hn03hn06hn07hn08,gmdn07} does 
have a chance to answer the above mentioned open questions. The question is, of course, 
whether and when the author~\footnote{I started the project named the {\it Approach 
unifying spins and charges} fifteen years ago, proving alone or together with collaborators  
step by step that such a theory has a real chance to answer the open questions 
of the {\it Standard model}. The names of the collaborators and students can be found in the 
cited papers. }
(and collaborators) will succeed to prove, that this theory really manifests in the low energy regime 
as the effective theory, postulated by the {\it Standard model}. 

Let me briefly present first  i.) the starting assumptions of the {\it Approach} (in this section), ii.)  
 a brief explanation of what does it offer (section~\ref{predictions}), iii.) the so far made  
proofs and the obtained positive results (section~\ref{LHCdarkmatter})  and at the end the open problems 
which are studying or are  waiting to be studied (section~\ref{kkandotherproblems}).

The {\it Approach} assumes~\cite{n92n93n95n01n07,pn06,hn00hn02hn03hn06hn07hn08,gmdn07} 
a simple action in $d=(1+13)$-dimensional space  
\begin{eqnarray}
S &=& \int \; d^dx \;{\mathcal L}_{f} +  \int \; d^dx \;  {\mathcal L}_{g} 
\label{action}
\end{eqnarray}
with the Lagrange density  for a spinor, which carries  in $d=(1+13)$ {\it two  kinds of the spin} and {\it 
no charges}. One kind of spin is {\it the Dirac spinor} in $(1+(d-1))$-dimensional space which, 
like in the Kaluza-Klein-like theories, takes care of the spin in $d=(1+3)$ and all the charges, 
and the {\it second kind}, the new one,  which takes care of the families~\footnote{This is the 
only theory in the literature to my knowledge, which does not explain the appearance of families 
by just postulating their numbers on one or another way, through the choice of a group, for example, 
but by offering the mechanism for generating families.}. There exist only two kinds of spins, 
corresponding to the left and the right multiplication of the Clifford algebra objects. 
The two kinds of spins are represented 
by the two kinds of the Clifford algebra objects~\cite{n92n93n95n01n07,pn06,hn00hn02hn03hn06hn07hn08,gmdn07} 
\begin{eqnarray}
\label{clifford}
S^{ab}&=& \frac{i}{4} (\gamma^a \gamma^b - \gamma^b \gamma^a),\quad   
\tilde{S}^{ab} = \frac{i}{4} (\tilde{\gamma}^a \tilde{\gamma}^b - \tilde{\gamma}^b \tilde{\gamma}^a), \nonumber\\
\{\gamma^a, \gamma^b \}_{+}&=& 2 \eta^{ab}= \{\tilde{\gamma}^a, \tilde{\gamma}^b \}_{+}, \quad 
 \{\gamma^a, \tilde{\gamma}^b \}_{+}=0,\quad \{S^{ab},\tilde{S}^{cd}\}_{-}= 0. 
\end{eqnarray}
The spinor interacts correspondingly only with the vielbeins $f^{\alpha}{}_{a}$ and the two kinds of 
the spin connection fields, $ \omega_{ab\alpha}$ and  $\tilde{\omega}_{ab\alpha}$, 
\begin{eqnarray}
     {\mathcal L}_f &=& \frac{1}{2} (E\bar{\psi} \, \gamma^a p_{0a} \psi) + h.c. \nonumber\\
p_{0a }&=& f^{\alpha}{}_a \, (p_{\alpha}  - 
                     \frac{1}{2}  S^{ab}   \omega_{ab\alpha} - 
                    \frac{1}{2}   \tilde{S}^{ab}   \tilde{\omega}_{ab\alpha}) + 
                    \frac{1}{2E}\, \{p_{\alpha}, f^{\alpha}{}_{a} E \}.
\label{lspinor}
\end{eqnarray}
Correspondingly there is the Lagrange density for the gauge fields,  assumed to be 
linear in the curvature 
\begin{eqnarray}
\label{lgauge}
{\mathcal L}_{g} &= &E  \;(\alpha \, R + \tilde{\alpha}  \tilde{R}), \nonumber\\
R &=& f^{\alpha [a}f^{\beta b]} \;(\omega_{a b \alpha,\beta} - \omega_{c a \alpha}
\omega^{c}{}_{b \beta}),
\tilde{R} = f^{\alpha [a} f^{\beta b]} \;(\tilde{\omega}_{a b \alpha,\beta} - \tilde{\omega}_{c a \alpha}
\tilde{\omega}^{c}{}_{b \beta}) 
\end{eqnarray}
and it is also the torsion field $ {\cal T}^{\beta}{}_{ab}= f^{\alpha}{}_{[a} (f^{\beta}{}_{b]})_{, \alpha}
+ \omega_{[a}{}^{c}{}_{b]} 
f^{\beta}_{c}$. Indices $a$ and $\alpha $ define the index 
in the tangent space and the Einstein index, respectively, $m$ and $\mu $ determine the 
"observed" $(1+3)$-dimensional space, $s$ and $\sigma$ the indices of "non observed" dimensions.

The action~Eq.(\ref{action}), which manifests the families (through $\tilde{S}^{ab}$) as the 
equivalent representations to the 
spinor representation of $S^{ab}$ (as can be seen in the last term of Eq.~(\ref{clifford})), 
has a real chance to manifest after the spontaneous breaks 
of symmetries in the low energy regime,  as I shall comment bellow 
(sections~\ref{predictions},~\ref{LHCdarkmatter}), 
all the properties required 
by the {\it Standard model} in order to fit the so far observed data. The {\it Approach } does not only 
explain, for example, why  the left handed spinors carry the weak charge while the 
right ones do not, and predicts that for the spontaneous breaks of symmetries 
the vielbeins together with both kinds of the spin connection fields 
are responsible, as they  are 
also for the Yukawa couplings and the appearance of the scalar fields, but explains the mechanism 
for the appearance of families, predicting the fourth family to be possibly seen at the 
LHC (or at somewhat higher energies)~\cite{pn06,gmdn07,nB09}
and the stable fifth family, whose neutrinos and baryons with masses several hundred TeV/$c^2$ 
 might explain the appearance of the dark matter~\cite{gn09}.

The  {\it Approach} confronted and still confronts several problems 
(among them are the problems common to all 
the Kaluza-Klein-like theories which  have vielbeins and spin connections  
as the only gauge fields, without  additional gauge fields in the bulk), 
which we are studying step by step when searching for possible 
ways of spontaneous breaking of the starting symmetries,  leading to  
the properties of the observed families of fermions and gauge and scalar 
fields, and looking for predictions the  {\it Approach} is 
offering~\cite{hn00hn02hn03hn06hn07hn08,nB09,gn09}.

I kindly ask the reader to look for more detailed presentation of the {\it Approach} in the  
refs.~\cite{pn06,gmdn07,nB09} and the references therein. 
%
%
\section{The Low Energy Limit of the {\it Approach Unifying Spin and Charges}}
\label{predictions}
The action of Eq.~(\ref{action}) starts with the massless spinor, let say, of 
the left handedness. The spinor interacts with the vielbeins and through two kinds of  
spins ($S^{ab}$ and $\tilde{S}^{ab}$) with the two kinds of the spin connection fields.
It was shown~\cite{n92n93n95n01n07,pn06,hn00hn02hn03hn06hn07hn08,gmdn07} that the Dirac 
kind of the Clifford algebra objects ($\gamma^a$) determines, when the group 
$SO(1,13)$ is analysed with respect to the {\it Standard model} groups in 
$d=(1+3)$ the spin and all the so far observed charges, manifesting the left handed quarks and leptons  
carrying the weak charge and the right handed weak chargeless quarks and leptons~\cite{pn06,gmdn07,nB09}. 
Accordingly  the Lagrange density ${\mathcal L}_f$ (Eq.~\ref{action},\ref{lspinor}) manifests after the 
appropriate breaks of the symmetries all the properties of one family of fermions as assumed by 
the {\it Standard model}: The three kinds of  charges coupling  fermions to the 
corresponding gauge fields, as presented bellow 
in the first term  on the right hand side of Eq.(\ref{lsmyukawa}).

The second kind ($\tilde{\gamma}^a$) of the 
Clifford algebra objects (defining the equivalent representations with respect to the Dirac one) 
determines families. Accordingly manifests the spinor Lagrange density, after the breaks 
of the starting symmetry ($SO(1,13)$ into $SO(1,7)\times U(1)\times SU(3)$ and further    
into $SO(1,3) \times SU(2) \times SU(2) \times U(1) \times SU(3)$)
the {\it Standard model-like} Lagrange density for massless spinors of  eight (four $+$ four) families 
(defined by $2^{8/2-1}=8$ spinor states for each member of one family). In the first  
successive break, appearing at around  $10^{13}$ GeV (or at a little lower scale),  
the upper four families (in the 
Yukawa couplings decoupled from the lower four families) obtain masses, while the lower four families 
stay massless (due to the fact that they are singlets with respect to the generators 
$\tilde{c}^{Ai}{}_{ab}\, \tilde{S}^{ab}$, forming one of the two  $SU(2) $ groups) and mass protected 
(since only the left handed spinors carry the weak charge).  
The Yukawa couplings are  presented  in the second term on the right 
hand side of Eq.(\ref{lsmyukawa}) bellow. The third term $("{\rm the \;rest}")$ in Eq.(\ref{lsmyukawa}) 
can still contribute to the second break. 
\begin{eqnarray}
{\mathcal L}_f &=& \bar{\psi}\gamma^{m} (p_{m}- \sum_{A,i}\; g^{A}\tau^{Ai} A^{Ai}_{m}) \psi 
+  \sum_{s=7,8}\;  \bar{\psi} \gamma^{s} p_{0s} \; \psi   +  {\rm the \;rest}. 
\label{lsmyukawa}
\end{eqnarray}
Here $\tau^{Ai}$ ($= \sum_{a,b} \;c^{Ai}{ }_{ab} \; S^{ab}$) determine the hyper ($A=1$), the weak ($A=2$) 
and the colour ($A=3$) charge: $\{\tau^{Ai}, \tau^{Bj}\}_- = i \delta^{AB} f^{Aijk} \tau^{Ak}$, $f^{1ijk}=0$, 
$f^{2ijk}=\varepsilon^{ijk}$, $f^{3ijk}$ is the $SU(3)$ structure tensor. 
In the final break (leading to 
$SO(1,3)   \times U(1) \times SU(3)$, which is the {\it Standard model} like break) the last four families 
obtain masses. 

The two breaks appear at two very different scales and are caused by vielbeins and both kinds 
of the spin connection fields in any of these cases, manifesting the two kinds of the Yukawa 
couplings and the two kinds of  scalar fields. The last break influences also the masses 
and the mixing matrices of the upper four families 
(in the Yukawa couplings decoupled from the lower ones) leading 
to the same gauge fields for all of the eight families~\cite{n92n93n95n01n07,pn06}.

Correspondingly manifests  
the ${\mathcal L}_{g} $ at observable energies all the three known gauge fields in the 
Kaluza-Klein-like way   
 \begin{eqnarray}
 \label{higgsgauge}
  e^{a}{}_{\alpha} = 
 \begin{pmatrix} \delta^{m}{}_{\mu}  & e^{m}{}_{\sigma}=0  \\
  e^{s}{}_{\mu}= e^{s}{}_{\sigma} E^{\sigma}{}_{Ai} A^{Ai}_{\mu} & e^s{}_{\sigma} \end{pmatrix}.
    \end{eqnarray}
%
Here $E^{\sigma Ai} = \tau^{Ai} \, x^{\sigma}.$
The scalar fields  and the gauge fields manifest 
through the vielbeins.
 
The {\it Approach} predicts two stable families; the first and the fifth. 
The quarks with the lowest mass 
among the upper four families, clustered into  fifth family baryons, are, together with the fifth family neutrinos,  
the candidates to  form the dark matter. For detailed calculations the reader can look at the 
ref.~\cite{arXiv09124532p119-135}. 
\section{Rough Estimations and Predictions}
\label{LHCdarkmatter}
Although we can not (yet, since it needs a lot of additional understanding and calculations) tell 
how do these spontaneous  breaks occur and at what energies do they 
occur, we still can make some estimations. At the break, which occurs at around $10^{13}$ GeV and 
makes the upper four families massive, the lightest (the stable one) of this four families  
obtains the mass, which is higher than several hundreds GeV  (to be in agreement with the 
experimental data and with the prediction of the {\it Approach} that there are four rather than three 
families with nonzero mixing matrices) and pretty  lower than the scale of the break. It is also 
meaningful to assume that the second break, which appears at much lower (that is weak scale) 
will not influence the masses of the upper four families considerably. 
Expressing the term $\sum_{s=7,8}\;  \bar{\psi} \gamma^{s} p_{0s} \; \psi$ of Eq.~(\ref{lsmyukawa})
in terms of the superposition of the $\omega_{st \sigma}$ and $\tilde{\omega}_{st \sigma}$ as 
dictated by the breaks, one ends up with the mass matrices first for the upper and, 
after the second break, for the lower four families~\cite{pn06,gmdn07} on the tree level. 
Since the $\omega_{st \sigma}$ and $\tilde{\omega}_{st \sigma}$ are not known, we only see the 
symmetries of the mass matrices. 

In the  subsection~(\ref{LHC}) I  present results of the estimations on the tree level 
of properties of the lower four families~\cite{pn06,gmdn07}, when the Yukawa couplings from 
Eq.~(\ref{lsmyukawa}) were taken as parameter and fited to the existing experimental data. 
Although we were not able to calculate the masses of the fourth family members, we calculated, 
for the assumed masses the mixing matrices, predicting that the fourth family might be seen 
at the LHC or at somewhat higher energies.
In the subsection~(\ref{darkmatter})  I present how do the experimental data--the 
cosmological ones and the data from direct measurements--limit the properties of the 
fifth family quarks: The present dark matter density limits (under the assumption that 
the fifth family baryons are mostly responsible for it) their masses in the interval  
$10 {\rm TeV} < m_{q_5}\, c^2 <  {\rm a \,  few\, hundreds \, TeV}$,
while the direct measurements, if they measure our fifth family baryons, predict the masses  
$m_{q_5} \ge 200  \frac{{\rm TeV}}{c^2}$.
These estimations will  
serve as a guide to  next more demanding studies of the predictions of the {\it Approach}. 
\subsection{Predicting  the fourth family properties}
\label{LHC}
After the last break (caused again by the vielbeins and the two kinds of the spin connection fields, 
connected now with the second $SU(2)$ symmetry break in the $\tilde{S}^{ab}$ sector)  
the action (Eq.~\ref{lsmyukawa}) manifests the properties postulated by the {\it Standard model}: 
the massive gauge fields $Z_\mu$ and $W^{\pm}_{\mu}$, the massless $U(1)$ and $SU(3)$ fields,  
the scalar filed (the {\it Standard model} Higgs) and the  mass matrices of quarks and leptons of the lower 
four families. 

It is a very difficult study to estimate the breaks and correspondingly the numerical values 
for parameters  which the {\it Standard} model fits from the experimental data. We are still  able 
to tell something about the mass matrices, if taking into account 
the symmetries of them, which we evaluated  on the tree level. The parameterised 
mass matrix is presented bellow~\cite{pn06,gmdn07}.
\begin{equation}
\label{massmatrices}
\left(
\begin{array}{cccc}
a_{\pm} & b_{\pm}           & -c_{\pm}          & 0 \\
b_{\pm} & a_{\pm}+ d_{1\pm} & 0                 & -c_{\pm}  \\
c_{\pm} & 0                 & a_{\pm}+ d_{2\pm} & b_{\pm} \\
0       & c_{\pm}           & b_{\pm}           & a_{\pm}+ d_{3\pm} \\
\end{array}
\right) 
\end{equation}
The parameters $a_{\pm}, b_{\pm},c_{\pm}$ distinguish among $u_i$ and $d_i$, or $\nu_i$ and $e_i$ 
($+$ for $u_i$ and $\nu_i$,  and $-$ for $d_i$ and $e_i$), while 
the parameters $d_{i\pm}$ distinguish among all the members of one family. 
Assuming that the contributions when 
going bellow the tree level would change the values and not the symmetries, and not paying 
attention on the $CP$ non conservation (studies of the discrete symmetries of the {\it Approach} are 
under consideration), that is assuming that the 
parameters are real, we fit with the Monte Carlo 
program these parameters to the existing data for the three families within the known accuracy. 
We did this for quarks and leptons. 
We were not able to tell the masses of the fourth family members unless requiring an additional 
symmetry~\cite{gmdn07}.  
For {\it the chosen values for the fourth family masses} (we took ($215, 285, 85, 170$ GeV/$c^2$ for the 
$u_4, d_4,\nu_4,e_4$ respectively), we were able to evaluate the mixing matrices, which fit the 
experimental data~\cite{gmdn07}. We present bellow, as an example, the 
mass matrix for the $u$-quarks 
%
\begin{eqnarray}
\label{mu}
\left(
\begin{array}{cccc}
(9  , 22) & (-150  , -83) & 0 & (-306  , 304) \\
(-150  , -83) & (1211  , 1245) & (-306  , 304) & 0 \\
0 & (-306  , 304) & (171600  , 176400) & (-150  , -83) \\
(-306  , 304) & 0 & (-150  , -83) & 200000\\ \nonumber
\end{array}
\right)
\end{eqnarray}
and the mixing matrix for the quarks
\begin{eqnarray}
\label{resultckmM}
 \left(\begin{array}{cccc}
	-0.974&-0.226&-0.00412&0.00218 \\
	0.226&-0.973&-0.0421&-0.000207 \\
	0.0055&-0.0419&0.999&0.00294 \\
	0.00215&0.000414&-0.00293&0.999 
 \end{array}
                \right).\nonumber
 \end{eqnarray}
We found the following masses of the four family members: 
$m_{u_i}/\textrm{GeV} = (0.0034, 1.15, 176.5, 285.2),\quad m_{d_i}/\textrm{GeV} = (0.0046, 0.11, 4.4, 224.0)$
and $m_{\nu_i}/GeV = ( 1\cdot 10^{-12}, 1\cdot 10^{-11}, 5\cdot
10^{-11},  84.0 ), m_{e_i}/GeV = (0.0005,0.106,1.8, 169.2)$.  
We are now studying the properties of the Yukawa couplings for the upper four and the lower four 
families bellow the tree level.  

Although we can not tell the masses of the fourth family yet, we can predict that the 
fourth family will sooner or later be experimentally confirmed, and due to my understanding, 
at much lower energies than the supersymmetric partners. 
\subsection{The fifth family baryons and neutrinos forming the dark matter}
\label{darkmatter}
%
%
The {\it Approach} predicts two times four families bellow the energy scale of $10^{13}$ GeV. 
The upper four families obtain masses when one of the two $SU(2)$ symmetries in both 
sectors--$S^{ab}$ and $\tilde{S}^{ab}$--breaks. The break of the second $SU(2)$ symmetry influences 
the  properties of the upper four families as well, so that they manifest the same charges and interact 
with the same gauge fields as the lower four families. It (slightly) influences also their masses. 

The lowest of the four upper families is stable (with respect to the age of the universe) and 
is accordingly the candidate to answer the open question of both standard models--the electroweak and the 
cosmological--that is about the origin of the dark matter in the universe. We have much less data about 
the upper four families than about the lower four families. The only data about the upper four families 
follow from the known properties of the dark matter.

In what follows I briefly present the estimations of the main properties of the fifth family members, 
making the assumption that the fifth family neutron is the lightest fifth family baryon and that 
the neutrino is the lightest fifth family lepton~\footnote{Let me mention as one of the arguments for 
the assumption that the neutron is the lightest baryon. While the electrostatic repulsion energy 
contributes around $1$ MeV more to the mass of the first family proton that it does to the 
first family neutron mass, this repulsion energy is for the fifth family proton, which is made out of 
quarks of the masses of several hundreds TeV/$c^2$, $100$ GeV(which is still only 1 per mil of 
the whole mass).}. Other possibilities are under consideration.
For known masses of quarks and leptons of the fifth family members and for known 
phase transitions in the universe when expanding and cooling down, the behaviour  of 
the fifth family members in the evolution of the universe as well as   when the clusters 
of galaxies are formed and when they interact with the ordinary  
matter follow (although the calculations are extremely demanded). We should know also 
the fifth family baryon/antibaryon asymmetry. We have not yet studied 
the masses and the mixing matrices for the upper four families and not yet the baryon/antibaryon 
asymmetry within the {\it Approach}. From what it is presented in subsection~\ref{LHC}  
it follows that the masses of the fifth family members are above 1 TeV/$c^2$, since they are expected 
to be  above the fourth family masses.  
 
 We assume in what follows no baryon/antibaryon 
asymmetry, which, as we shall se, does  not influence much the results  for large enough masses 
(above several tens TeV/$c^2$) of the fifth 
family quarks. For high enough masses the one gluon exchange  determines the properties of quarks in 
baryons as well as the quarks' interaction with the cosmic plasma when the temperature is above 
$1$ GeV/$k_b$ and correspondingly their freezing out of the plasma  
and their forming the fifth family neutrons during the expansion~\cite{gn09}.

The fifth family baryons and antibaryons, decoupling from the plasma mostly before  or during the 
phase transition,  
interact among themselves through the "fifth family nuclear force", which is for the masses of the 
fifth family quarks of the order of several hundred TeV/$c^2$ for a factor  $10^{-10}$ smaller than the 
ordinary (first family) nuclear force~\cite{gn09}.
Correspondingly the fifth family baryons and 
antibaryons interact 
in the dark matter clouds in the galaxies and among the galaxies, when they scatter, dominantly 
with the weak force. With the ordinary matter they interact through the "fifth family nuclear force" 
and are obviously not  WIMPS (weakly interacting massive particles, if WIMPS are meant to interact 
only with the weak force). 

We estimate  that the fifth family neutrinos with masses above TeV/$c^2$ and bellow $200$ TeV/$c^2$ 
contribute to the dark matter and to the direct measurements less than the fifth family 
neutrons~\footnote{Both, quarks and leptons of the fifth family (as well as antiquarks and 
antileptons), experience the electroweak break, having 
accordingly different electroweak properties before and after the break. We are studying 
these properties. The fifth family quarks, 
which do not decouple from the plasma before the colour phase transition, which starts 
bellow $1$ GeV/$c^2$, are influenced by the colour phase transition as well. We estimate~\cite{gn09} 
that in the colour phase transition all coloured quarks and antiquarks either annihilate or form 
colourless clusters and decouple out of plasma}. 
 \subsubsection{ The fifth family quarks in the expanding universe}
  To solve the coupled Boltzmann equations for
 the number density of the fifth family quarks and the colourless clusters of the quarks in the plasma 
 of all the other fermions (quarks, leptons) and bosons (gauge fields) in the thermal equilibrium  in the 
 expanding universe we estimate~\cite{gn09} the cross sections for the annihilation of quarks with 
 antiquarks and for forming clusters. We do this within some uncertainty intervals, which take 
 into account the roughness of our estimations. Knowing only the interval for possible values of masses 
 of the fifth family members we solve the Boltzmann equations  for several values of quark masses, 
 following the decoupling of the fifth family quarks and the fifth family neutrons out of the plasma 
 down to the temperature $1$ GeV/$k_b$ when the colour phase transition starts. 
 
 The fifth family neutrons and antineutrons, packed into very tiny clusters so that they are 
 totally decoupled  from the plasma, do not feel the colour phase transition, while the fifth family quarks and 
 coloured clusters (and antiquarks) do. Their scattering cross sections grow due to the 
 nonperturbative behaviour of  gluons (as do the scattering cross sections of all the other 
 quarks and antiquarks). While the three of the lowest four families decay into the first family
 quarks, due to the corresponding Yukawa couplings, the fifth family quarks can not. Having the 
 binding energy a few orders of magnitude larger than $1$ GeV and moving in the rest of plasma of 
 the first family quarks and antiquarks and of the other three families and gluons as very 
 heavy objects with large scattering cross section, the fifth family coloured objects annihilate with their 
 partners or form the colourless clusters (which result in the decoupling from the plasma) long before 
 the temperature falls bellow a few MeV/$k_b$ when the first family quarks start to form the bound states. 
 
 Following further the fifth family quarks in the expanding universe up to today and 
 equating the today's dark matter density with the calculated one, we estimated  the mass 
 interval of the fifth family quarks to be
 %
($10\; {\rm TeV} < m_{q_5}\, c^2 <  {\rm a \,  few\, hundreds \, TeV}$).
%
The detailed calculations with all the needed explanations can be found in ref.~\cite{gn09}. 
\subsubsection{Dynamics of a heavy  family baryons in our galaxy and the direct measurements}
Although the average properties of the dark matter in the Milky way are pretty well known 
(the average dark matter density at the position of our Sun is expected to be 
$\rho_0 \approx 0.3 \,{\rm GeV} /(c^2 \,{\rm cm}^3)$,  
and the average velocity of the dark matter constituents around the centre of our 
galaxy is expected to be approximately velocity of our  Sun), their real local properties are 
known much less accurate,  within the factor of $10$~\footnote{In a simple model that all 
the clusters at any radius $r$ from the centre 
of our galaxy travel in all possible circles around the centre so that the paths are 
spherically symmetrically distributed, the velocity of a cluster at the position of 
the Earth is equal to $v_{S}$, the velocity of our Sun in the absolute value,
but has all possible orientations perpendicular to the radius $r$ with  equal probability.
In the model 
that the clusters only oscillate through the centre of the galaxy, 
the velocities of the dark matter clusters at the Earth position have values from 
zero to the escape velocity, each one weighted so that all the contributions give  
$ \rho_{dm} $.
}. 
 When evaluating~\cite{gn09} the number of events which our fifth family members 
 (or any stable heavy family baryons or neutrinos) trigger in the direct 
 measurements of DAMA~\cite{rita0708} and CDMS~\cite{cdms} experiments, we take into account 
 all these uncertainties as well as the uncertainties in the theoretical estimation and the 
 experimental treatments. Let the dark matter member hits the Earth with the  velocity 
 $\vec{v}_{dm \,i}$. The velocity of the Earth around the centre of the galaxy is equal to:  
$\vec{v}_{E}= \vec{v}_{S} + \vec{v}_{ES} $, with $v_{ES}= 30$ km/s and 
$\frac{\vec{v}_{S}\cdot \vec{v}_{ES}}{v_S v_{ES}}\approx \cos \theta, \theta = 60^0$, 
$v_{S}=(100-270)$ km/s. 
The dark matter cluster of the $i$- th  velocity class  hits the Earth with the velocity:  
$\vec{v}_{dmE\,i}= \vec{v}_{dm\,i} - \vec{v}_{E}$. 
Then the flux of our dark matter clusters hitting the Earth is:    
$\Phi_{dm} = \sum_i \,\frac{\rho_{dm i}}{m_{c_5}}  \,
|\vec{v}_{dm \,i} - \vec{v}_{E}|  $, which  can be approximated 
by $\Phi_{dm} = \frac{\rho_0 \,\varepsilon_{\rho} }{m_{c_5}}\,\{ \varepsilon_{v_{dmS}} 
\, v_S \,  + $ 
$ \varepsilon_{v_{dmES}}\, v_{ES} \cos \theta \, \sin \omega t \}$.  
The last term determines the annual modulations observed by DAMA~\cite{rita0708}.
We estimate (due to experimental data and our theoretical evaluations) that 
$\frac{1}{3} < \varepsilon_{v_{dmS}} < 
3$ and $\frac{1}{3} < \frac{\varepsilon_{v_{dmES}}}{\varepsilon_{v_{dmS}}} < 3$. 

The cross section for our fifth family baryon to elastically
scatter  on an ordinary nucleus with $A$ nucleons is 
$\sigma_{A} \approx  
\frac{1}{\pi \hbar^2} <|M_{c_5 A}|>^2 \, m_{A}^2$,  
where  $m_{A}  $  is the mass of the ordinary nucleus~\footnote{Although our fifth family 
neutron or antineutron is very tiny, of the size of  $10^{-5}$ fm, it is very massive (a few hundreds 
TeV/$c^2$). When a  quark of the ordinary nucleus scatter on it,  the whole nucleus scatters with it, 
since at the recoil energies the quarks are strongly bound. }. 
Since  scattering is expected to be coherent 
the cross section is  almost independent of the recoil 
velocity of the nucleus. Accordingly is 
the cross section   
$\sigma(A) \approx \sigma_{0} \, A^4 \, \varepsilon_{\sigma}$, with 
$\sigma_{0}\, \varepsilon_{\sigma}$, which is $9\,\pi r_{c_5}^2 \, 
\varepsilon_{\sigma_{nucl}} $, with 
 $\frac{1}{30} < \varepsilon_{\sigma_{nucl}} < 30$ (taking into account the roughness 
with which we treat our  heavy baryon's properties and the scattering procedure) 
when the ''nuclear force'' dominates. 
In all the expressions the index $c_5$ denotes the fifth family cluster, while the index $nucl$ 
denotes the ordinary nucleus. 

Then the number of events per second  ($R_A$) taking place 
in $N_A$ nuclei of some experiment  is  due to the flux $\Phi_{dm}$ 
equal to $
R_A \,\varepsilon_{cut}= \varepsilon_{cut} \, N_A \,  \frac{\rho_{0}}{m_{c_5}} \;
\sigma_0 \,A^4 \, v_S \, \varepsilon \, ( 1 + 
\frac{\varepsilon_{v_{dmES}}}{\varepsilon_{v_{dmS}}} \, \frac{v_{ES}}{v_S}\, \cos \theta
\, \sin \omega t), $ where we estimate  that  $\frac{1}{300} < \varepsilon < 300$     
demonstrates  the uncertainties in the knowledge about the dark matter dynamics 
in our galaxy and our approximate treating of the dark matter properties, 
while  $\varepsilon_{cut}$ determines the uncertainties in the detections.   

Taking these evaluations into account we predict that if DAMA~\cite{rita0708} is  measuring 
our fifth family (any heavy stable) baryons then CDMS~\cite{cdms} (or some other experiment) will 
measure in a few years these events as well~\cite{gn09} provided that 
our fifth family quarks masses are higher than 
%
$m_{q_5} \ge 200  \frac{{\rm TeV}}{c^2}$.
%
%
%
\section{Concluding Remarks}
\label{conclusion}
I demonstrated in my talk that the {\it Approach unifying spin and charges}, which  
assumes in $d=(1+13)-$dimensional space a  simple action for a gravitational 
field (manifesting with the spin connections and the vielbeins) and massless 
fermions carrying only {\it two kinds of the spin} (the one presented by  
the Dirac matrices and the additional one anticommuting with the Dirac one--there is no the 
third kind of the spin), no charges,  
shows a new way beyond the {\it Standard model of the electroweak 
and colour interactions}. It namely, for example, explains the origin of families, of  
the Yukawa couplings, of the charges, the appearing of the corresponding gauge fields 
and of the (two kinds of) scalar fields, it explains   
why only the left handed spinors carry the weak charge while the right handed ones do not, 
where does the dark matter originate, and others. 
The action  manifests at low (observable) energies after particular breaks 
of symmetries two times four families, with no Yukawa couplings (in comparison with 
the age of the universe) among the lower and the upper four families.

The {\it Approach} predicts  the fourth family, to be possibly observed at the LHC 
or at somewhat higher energies and the stable fifth family, whose baryons and neutrinos  form 
the dark matter. I discussed briefly the properties of the lower four families, estimated at 
the tree level.
Following the history of the fifth family members in the expanding universe up to today and 
estimating also the scattering properties of this fifth family on the ordinary matter, the evaluated 
masses of the fifth family quarks,  under the assumption that the lowest mass fifth family baryon 
is the fifth family neutron, are in the interval 
\begin{eqnarray}
\label{all}
200 \,{\rm TeV} < m_{q_5}\,c^2 <  10^5 \,{\rm TeV}.
\end{eqnarray} 
The fifth family neutrino mass is estimated to be in the interval: 
a few TeV $< m_{\nu_5}\,c^2 <$ a few hundreds TeV. 
\subsection{Problems to be solved}
\label{kkandotherproblems}
In the sections~(\ref{conclusion},\ref{introduction}) the promising sides of the 
{\it Approach unifying spin and charges} are briefly over viewed and in the rest of talk 
the achievements presented. 
One of the conclusions we can make is that--since there are two kinds of the 
Clifford algebra objects and not only one (successfully used by Dirac 80 years ago)  
 describing the spin of fermions and in the {\it Approach}  the spin and the charges, 
and since the other generates the equivalent representations with respect to the first one, while
the families are the equivalent representations with respect to the spin and the
charges--the second kind of the spin can be used to describe families, or even must, since 
if not, we should explain,
why only one kind of the spin manifests at the low energy regime.

There are, however, several questions which should be solved  before 
accepting the {\it Approach} as the right way beyond the {\it Standard model},  
%
offering the  right explanation  for the origin of families, charges and gauge fields.  
Since the {\it Approach} assumes that the dimension 
of the space time is larger than (1+3), like do the Kaluza-Klein-like theories as well as  
the theories with strings and membranes, it must be asked what is at all the dimension of the space-time. 
What does cause or trigger the spontaneous breaks of symmetries in the evolution of the universe? 
What does determine the phase transitions? 

We put a lot of efforts in understanding the properties of higher dimensional spaces and in 
understanding the properties of fermions after breaking symmetries. 
The reader can find more about our understanding and the proposed solutions in the 
references~\cite{hn00hn02hn03hn06hn07hn08}. 
%
%
It is  hard to evaluate how do the spontaneous (non adiabatic) breaks occur, what causes them and 
how do they influence 
the properties of all kinds of fields. To understand better the differences in the properties of a family members, 
we are estimating their properties bellow the tree level. Is it the expectation that 
there are effects  bellow the tree 
level which are responsible for the differences in properties of family members (although some 
differences manifest already on the tree level) correct? Will such calculations show that 
the fifth family $u$-quark is heavier than the $d$-quark  mainly due to 
 the repulsive electrostatic contribution ($\approx 100$ GeV)? 

The estimation of the behaviour of the coloured fifth family clusters during 
the colour phase transition  leads to the conclusion that the fifth family quarks 
either annihilate or form the fifth family neutrons or antineutrons. Will 
more accurate evaluations confirm this estimation? How many of the fifth family 
 neutrinos and neutrons annihilate in the  interval bellow the weak phase transition  
 due to possibly large weak cross section 
 for the annihilation in this region? 
 
 
 Many a question presented here are under consideration, many of them not even written here wait to be studied. 
 I invite the audience to contribute to next steps of (to my understanding) a really promising 
 way beyond the {\it Standard model}. 
%
%
 %

\end{document}